\documentclass[aps,prl,showpacs,twocolumn]{revtex4}
\usepackage{amsmath}
\usepackage{amsmath}
\usepackage{graphicx}
\usepackage{epsfig}
\usepackage{color}
\usepackage[all]{xy}

\newcommand{\be}{\begin{equation}}
\newcommand{\ee}{\end{equation}}
\newcommand{\bea}{\begin{eqnarray}}
\newcommand{\eea}{\end{eqnarray}}
\newcommand{\bqa}{\begin{eqnarray}}
\newcommand{\eqa}{\end{eqnarray}}
\newcommand{\bqs}{\begin{eqnarray*}}
\newcommand{\eqs}{\end{eqnarray*}}
\newcommand{\beq}{\begin{equation}}
\newcommand{\eeq}{\end{equation}}

\begin{document}

\draft
\noindent{{\it submitted}, Physical Review Letters \\   
\title{ Symmetry Relations for Trajectories of a Brownian Motor}
\author{R. Dean Astumian}
\email{astumian@maine.edu}
\affiliation{University of Maine, Orono, Maine, USA}

\date{\today}

 \begin{abstract}
 A Brownian Motor is a nanoscale or molecular device that combines the effects of thermal noise, spatial or temporal  asymmetry, and directionless input energy to drive directed motion.     Because of the input energy, Brownian motors function away from thermodynamic equilibrium and concepts such as linear response theory, fluctuation dissipation relations, and detailed balance do not apply.  The {\em generalized} fluctuation-dissipation relation, however, states that even under strongly thermodynamically non-equilibrium conditions the ratio of the probability of a transition to the probability of the time-reverse of that transition is the exponential of the change in the internal energy of the system due to the transition.  Here, we derive an extension of the generalized fluctuation dissipation theorem for a Brownian motor  for the ratio between the probability for the motor to take a forward step and the probability to take a backward step.  
\end{abstract}

\maketitle
\vspace{-5pt}
A Brownian Motor is a nanoscale or molecular device that combines the effects of thermal noise, spatial or temporal  asymmetry, and directionless input energy to drive directed motion \cite{ast_sci97,rei_pr02,ast_pt02}.  Many biological  motile systems may be driven by Brownian motors \cite{ast_ebj98}, and chemists have been able to synthesize molecules that function as Brownian motors \cite{leigh_angchem07,fer_natnano06}.   In solution, viscous drag and thermal noise dominate the inertial forces that drive macroscopic machines.  Because of the strong viscous drag, the motion of such a Brownian motor is over-damped and in one dimension  can be described by the simple equation \cite{ons_pr53}
\be
R \dot{\alpha} - X = \epsilon(t)
\ee
where $\epsilon(t)$  is Gaussian noise with mean $\mu = 0$ and variance $\sigma^2 = 2 R k_B T/dt$, and R is the coefficient of viscous friction. In the following we use units where the thermal energy $k_B T = 1$.  The generalized force $X = X(\alpha,\psi(t))$  can be written as the gradient of a scalar potential  $X = -\partial H/\partial \alpha$ where
\be
H(\alpha, \psi(t)) = U(\alpha) + \psi(t) z(\alpha) 
\ee
 is the sum of an intrinsic potential due to chemical interactions and any external load and an external time dependent forcing term that is the product of canonically conjugate intensive and extensive thermodynamic parameters  $z(\alpha)$ and $\psi(t)$, respectively \cite{ast_pra89}.  The conjugate parameters include, e.g., molecular volume and pressure, entropy and temperature, or dipole moment and field.  The underlying system is typically spatially periodic (possibly with a homogeneous force or load $F$) so that $U(\alpha + L) = U(\alpha) + \Delta U$, where $\Delta U = F L$,  and $z(\alpha + L) = z(\alpha)$.  

For any fixed value of $\psi$ detailed balance requires 
\begin{equation}
\frac{P(\alpha_i + L, {\cal T}|\cdots|\alpha_i,0)}{P^\dagger(\alpha_i, {\cal T}|\cdots| \alpha_i + L,0)}
= e^{-\Delta U}.
\end{equation}
 where  $P(\alpha_i + L, {\cal T}|\cdots|\alpha_i,0)$ is the conditional probability density that a particle starting  at position $\alpha_i$ at time $0$ goes to position $\alpha_i + L$ at time ${\cal T}$ by the specific trajectory (sequence of positions and times) denoted by $\cdots$, and $P^\dagger(\alpha_i, {\cal T}|\cdots| \alpha_i + L,0)$ is the conditional probability to follow the reverse of that process. The ratio depends only on the difference in energy between the initial and final points.   It further holds that
 \begin{equation}
\frac{P( L, {\cal T}|0,0)}{P(0, {\cal T}| L,0)} = e^{-\Delta U} 
\end{equation}
where the net probability $P( L, {\cal T}|0,0) = \int_0 \cdots \int^L P(\alpha_i + L, {\cal T}|\cdots|\alpha_i,0)$ is the integral over all trajectories from $(0,0)$ to $(L, {\cal T})$.

A time dependent modulation, $\psi(t)$,  causes dissipation and breaks detailed balance, in which case Eqs. (3) and (4) do not hold.  It is even possible to have 
\begin{equation}
\frac{P( L, {\cal T}|0,0)}{P(0, {\cal T}| L,0)} > 1; \,\,\,\, e^{-\Delta U} < 1
\end{equation}
where the external stimulus $\psi(t)$  provides energy to drive uphill motion \cite{ast_prl94,pro_prl94}. 

\begin{figure}[h]
\centerline{\includegraphics[scale=.40]{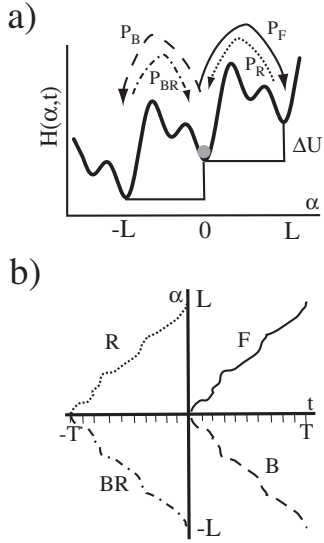}}
\caption{Depiction of symmetry related trajectories of a Brownian particle in a periodic ratchet potential.  a)  Snapshot of the potential described in Eq. (22) with a particle at $\alpha = 0$.  For {\em any} constant $\psi$, the forward (F) and backward reverse (BR) trajectories are identical (and $P_{\rm F} \equiv  P_{\rm BR}$), as are the time reverse (R) and backward (B) trajectories (and $P_{\rm R} \equiv  P_{\rm R}$) and by detailed balance $P_{\rm F}/P_{\rm B} = e^-\Delta U$ and the net motion of the particle is to the left.  External modulation  $\psi(t)$ breaks detailed balance and can drive net motion to the right against the load.  b)  With an external forcing $\psi(t)$ all four trajectories - F,R,B, BR - are distinct from one another.  As described in the text, an extension of the generalized fluctuation dissipation relation can be used to derive the ratios of the probability densities for these symmetry related trajectories. }
\end{figure}

The {\em generalized} fluctuation-dissipation theorem \cite{boch_jetp77,boch_physa81} states that even under strongly thermodynamically non-equilibrium conditions the ratio of the probability of a forward (F) transition to the probability of the {\em time-reverse} (R) of that transition is the exponential of the change in the internal energy of the system due to the transition
\begin{equation}
\frac{P_{\rm F}( L, {\cal T}|\cdots|0,0)}{P_{\rm R}(0, {\cal T}|\cdots|  L,0)}
= e^{W-\Delta U}.
\end{equation}
where W is the work supplied to the system by the external modulation in the forward trajectory.   Here, an extension of the generalized fluctuation dissipation theorem for a Brownian motor is derived  to obtain the ratio between the probability for the motor to take a forward step and the probability to take a backward step in {\em forward time}.

First, we write equation (1) as a more rigorous finite difference or update equation and convert to unit normal Gaussian noise $N(0,1)$ \cite{gil_ajp95} 
\be
\alpha_{\rm i+1} - \alpha_{\rm i} - R^{-1} X_{\rm  i+1}  \Delta t = \sqrt{2 R^{-1} \Delta t} \,\, N(0,1)
\ee
The time interval $\Delta t$ is chosen to be sufficiently short that the change in position $\Delta \alpha$ is very small.   We used the relation $N(\mu,\sigma^2) = \mu + \sigma N(0,1)$ where $N(0,1)$ is a Gaussian random variable with zero mean and unit variance the values, $n$, of which occur with probability, $P(n) = \exp{\left(-n^2/2\right)}/(\sqrt{2 \pi})$.  Any two values of $n$ are uncorrelated  $\left<n_{\rm i} n_{\rm k}\right>  = \delta_{\rm i,k}$.   

Broken symmetry is an essential feature of a Brownian motor, so we split each of the position dependent terms into even and odd components $U(\alpha) = U^{\rm e}(\alpha) + U^{\rm o}(\alpha)$ and $z(\alpha) =z^{\rm e}(\alpha) + z^{\rm o}(\alpha)$, where for any function $f^{\rm e}(-\alpha) = f^{\rm e}(\alpha)$ and $f^{\rm o}(-\alpha) = -f^{\rm o}(\alpha)$.  Finite difference expressions for the even and odd components of the generalized force, $X_{\rm  i+1} = X^{\rm e}_{\rm i+1} + X^{\rm o}_{\rm  i+1}$, are
\be
X^{\rm (e,o)}_{\rm  i+1} = -\frac{ \Delta U^{\rm (o,e)}_{\rm  i+1} + \psi_{\rm i+1}  \Delta z^{\rm (o,e)}_{\rm  i+1} }{\alpha_{\rm i+1} - \alpha_{\rm i}} 
\ee
where $\Delta f^{\rm k}_{\rm  i+1} = f^{\rm k}(\alpha_{\rm i+1}) - f^{\rm k}(\alpha_{\rm i})$ for $f = U,z$ and $k = e,o$.  For every forward trajectory $\{ \alpha(t), \psi(t) \}$ with probability $P_{\rm F}$, defined by 
\bea
{\rm F} \equiv	0 \stackrel{\psi_1}{\longrightarrow} 
		\alpha_1 \stackrel{\psi_2}{\longrightarrow} 
		\cdots \stackrel{\psi_{\rm m-1}}{\longrightarrow} 
		 \alpha_{\rm m-1}  \stackrel{\psi_{\rm m}}{\longrightarrow} 
		L, \nonumber \\
		\nonumber \\
P_{\rm F}  =  \prod_{\rm i=0}^{\rm M-1} P(\alpha_{\rm i+1}|\alpha_{\rm i}; \psi_{\rm i+1}),
\eea 
there are three symmetry related trajectories.   One is a time reverse trajectory \cite{crooks_jstatphys98} $\{ \alpha(-t), \psi(-t) \}$ obtained by switching the sign of time. For a time periodic system reversing time is equivalent to the transformation $t \rightarrow ({\cal T} - t)$ , 
\bea
{\rm R} \equiv	L \stackrel{\psi_{\rm m}}{\longrightarrow} 
		\alpha_{\rm m-1} \stackrel{\psi_{\rm m-1}}{\longrightarrow} 
		\cdots \stackrel{\psi_{\rm 2}}{\longrightarrow} 
		 \alpha_{\rm 1}  \stackrel{\psi_{\rm 1}}{\longrightarrow} 
		0, \nonumber \\
		\nonumber \\
		 P_{\rm R}  =  \prod_{\rm i = 0}^{\rm M-1} P(\alpha_{\rm i}|\alpha_{\rm i+1}; \psi_{\rm i+1}),
\eea 
Another is a backward trajectory $\{ -\alpha(t), \psi(t) \}$ obtained by switching the sign of the position variable.  For a space periodic system this is equivalent to the transformation $\alpha \rightarrow (L - \alpha)$, 
\bea
{\rm B} \equiv	0 \stackrel{\psi_1}{\longrightarrow} 
		-\alpha_1 \stackrel{\psi_2}{\longrightarrow} 
		\cdots \stackrel{\psi_{\rm m-1}}{\longrightarrow} 
		- \alpha_{\rm m-1}  \stackrel{\psi_{\rm m}}{\longrightarrow} 
		-L, \nonumber \\
		\nonumber \\
		 P_{\rm B}  =  \prod_{\rm i = 0}^{\rm M-1} P(-\alpha_{\rm i+1}|-\alpha_{\rm i}; \psi_{\rm i+1}) .
\eea  
 The third is a backward reverse trajectory $\{ -\alpha(t), \psi(-t) \}$ obtained by switching the sign of both time and of the position variable.  For a time and space periodic system this is equivalent to the transformation $\alpha \rightarrow (L - \alpha)$ and $t \rightarrow ({\cal T} - t)$, 
\bea
{\rm BR} \equiv	-L \stackrel{\psi_{\rm m}}{\longrightarrow} 
		-\alpha_{\rm m-1} \stackrel{\psi_{\rm m-1}}{\longrightarrow} 
		\cdots \stackrel{\psi_{\rm 2}}{\longrightarrow} 
		- \alpha_{\rm 1}  \stackrel{\psi_{\rm 1}}{\longrightarrow} 
		0, \nonumber \\
		\nonumber \\
		 P_{\rm BR}  =  \prod_{\rm i = 0}^{\rm M-1} P(-\alpha_{\rm i}|-\alpha_{\rm i+1}; \psi_{\rm i+1}),
\eea

Viewing Eq. (7) as a mapping between the ``noise" space and ``position" space \cite{bier_pla98,bier_pre99}, the conditional probability density given that the system is at position $\alpha_{\rm i}$ after the i$^{th}$ step, and that the value of the field is $\psi_{\rm i+1}$ for the (i+1)$^{st}$ step is seen to be
\be
 P(\alpha_{\rm i+1}|\alpha_{\rm i},\psi_{\rm i+1}) =  \frac{e^{\frac{-\left(\Delta \alpha - R^{-1} X_{\rm i+1}  \Delta t \right)^2}{4 R^{-1} \Delta t}}}{\sqrt{4 \pi R^{-1} \Delta t}}.
 \ee
where $\Delta \alpha =  (\alpha_{\rm i+1}-\alpha_{\rm i})$.  The ratio of the probability density for the forward and time reverse step is 
\be
 \frac{P(\alpha_{\rm i+1}|\alpha_{\rm i},\psi_{\rm i+1})}{P(\alpha_{\rm i}|\alpha_{\rm i+1},\psi_{\rm i+1})} =  e^{ X_{\rm  i+1}  \Delta \alpha }.
\ee
and the ratio between the forward and time reverse trajectory is 
\be
\frac{P_{\rm F}}{P_{\rm R}}=  \exp{\left(\displaystyle\sum_{\rm i=0}^{\rm M-1} X_{\rm  i+1}\Delta \alpha\right)} =  e^{W_{\rm F} -\Delta U }
\ee
where 
\be
W_{\rm F} = \sum_{\rm i=0}^{\rm M-1} \psi_{\rm i+1}\left(\Delta z^e(\alpha)+\Delta z^o(\alpha)\right)
\ee
is the total external work done in the forward trajectory.   Eq. (15) is the generalized fluctuation dissipation relation \cite{boch_jetp77,boch_physa81} and the change in the internal energy of the system, $\Delta E = \Delta U - W_{\rm F}$, is the dissipated work. The ratio of the probability density for a backward and backward time reverse is similarly obtained,
\be
\frac{P_{\rm B}}{P_{\rm BR}}  =  e^{W_{\rm B} + \Delta U }
\ee
where 
\be
W_{\rm B} = \sum_{\rm i=0}^{\rm M-1} \psi_{\rm i+1}\left(\Delta z^e(\alpha)-\Delta z^o(\alpha)\right)
\ee
is the total external work done in the backward trajectory.  Finally, the ratio between a forward and backward step is
\be
 \frac{P(\alpha_{\rm i+1}|\alpha_{\rm i},\psi_{\rm i+1})}{P(-\alpha_{\rm i + 1}|-\alpha_{\rm i},\psi_{\rm i+1})} =  e^{ X^{\rm e}_{\rm  i+1}  \Delta \alpha - X^{\rm e}_{\rm i+1}X^{\rm o}_{\rm i+1} R^{-1}  \Delta t }
  \ee
The central result of this paper, the ratio for the probability densities for a forward and backward trajectory, follows immediately
\be
\frac{P_{\rm F}}{P_{\rm B}}=  e^{- \Delta U} e^{-R^{-1} \int_0^{\cal T} X^{\rm e}X^{\rm o}  dt }
\ee
where we have taken the limit $\Delta t \rightarrow 0$ to get the integral form.  Equation (20) highlights the importance of broken symmetry - if either $X^{\rm o}$ or $X^{\rm e}$ is zero, the ratio of the probability for a forward step to a backward step is governed solely by the homogeneous force acting on the system and is independent of the work pumped in by the time dependent modulation.  Consider the standard ratchet potential below \cite{ast_pt02}
\bea
&U(\alpha) = U_0 \cos{(2 \alpha)} + F \alpha \nonumber \\
\\
&z(\alpha) =  z_0\left[ \cos{(s \pi)} \cos{( \alpha)} + \sin{(s \pi)} \sin{( \alpha)}\right] \nonumber
\eea 
where the asymmetry parameter is $-1 \leq s \leq 1$.   For $U_0 > \psi_0 z_0$, where $\psi_0$ is the amplitude of the external modulation, a single spatial period of the potential has two clearly defined energy wells, say A and B,  where the sign of $\psi$ determines the relative energies of the two wells, and the relative heights of the barriers between them.  We can then describe the motion as a random walk between the two wells,\xymatrix{
 A \ar @/^/ @{_{<}-^{>}} [r]
   \ar @/_/ @{_{<}-^{>}} [r] & B },  where a clockwise transition indicates a half-step to the right, and a counterclockwise transition indicates a half-step to the left.  The  transition constants are 
\bea   
&\roarrow{k}_{\rm AB} = k_0 \left[ e^{- \Delta U} (\phi^{\rm e} \phi^{\rm o}) \right]^{1/4} \nonumber  \\
&\roarrow{k}_{\rm BA} = k_0 \left[ e^{- \Delta U} /(\phi^{\rm e} \phi^{\rm o})\right]^{1/4} \nonumber \\
&\loarrow{k}_{\rm AB} = k_0 \left[ e^{ \Delta U} (\phi^{\rm e}/\phi^{\rm o}) \right]^{1/4} \nonumber \\
&\loarrow{k}_{\rm BA} = k_0 \left[ e^{ \Delta U} (\phi^{\rm o}/\phi^{\rm e}) \right]^{1/4}
\eea   
where $\phi^{\rm e} = e^{4 z_0 \psi(t) \cos{(s \pi)}}$ and $\phi^{\rm o} = e^{4 z_0 \psi(t) \sin{(s \pi)}}$.  Irrespective of the value of s or the form of $\psi(t)$ a corollary of detailed balance for rate processes,
\be
\frac{\roarrow{k}_{\rm AB} \roarrow{k}_{\rm BA} } {\loarrow{k}_{\rm AB}\loarrow{k}_{\rm BA} } =e^{-\Delta U}
\ee
holds at every instant.  The net motion can be solved analytically for small amplitude \cite{ast_jcp89} $\psi(t)$, and in special cases, such as square wave perturbation, for arbitrary amplitude \cite{ast_pra89} $\psi(t)$.  A similar model has been proposed for adiabatic transport \cite{sin_epl07}.   For the specific case that  $\psi(t)$ is externally generated dichotomic noise (\xymatrix{
  +\Psi  \ar@<1ex>[r]^\gamma
     & -\Psi \ar@<1ex>[l]^\gamma } ) in which $\psi(t)$ switches between $+\Psi$ and $-\Psi$ with a Poisson distributed random lifetime (average $1/\gamma$) ( a situation particularly relevant for Brownian motors that are driven, e.g., by the stochastic binding of chemical fuel molecule and release of product) the combined stepping/switching process can be described by a single diagram \cite{ast_pra89}
\be 
\xymatrix{
 A,+ \ar@{_{}-^{}}[d] \ar@/^1pc/ @{_{}-^{}} [r] \ar@/_1pc/ @{_{}-^{}} [r]& B,+ \ar@{_{}-^{}}[d]  \\
  A,- \ar@{_{}-^{}}[u] \ar@/^1pc/ @{_{}-^{}} [r] \ar@/_1pc/ @{_{}-^{}} [r]       & B,- \ar@{_{}-^{}}[u]     }
\ee
This picture emphasizes the idea of a minimal Brownian motor as two coupled two-state processes.   One process is the externally driven dichotomic modulation \xymatrix{
  +\Psi  \ar@<1ex>[r]^\gamma
     & -\Psi \ar@<1ex>[l]^\gamma } and the other is the thermally activated stepping  \xymatrix{
 A \ar @/^/ @{_{<}-^{>}} [r]
   \ar @/_/ @{_{<}-^{>}} [r] & B }.  
     
The overall diagram can be broken into six cycles \cite{ast_pra89} - two cycles for the uncoupled stepping, one with fixed $+\Psi$ and the other with fixed $-\Psi$, two cycles for the dissipative back and forth motion with no net stepping to the right or left, and two cycles describing net stepping coupled to the external fluctuation.  The last two, coupled, cycles are of most interest.  The forward, reverse, backward, and backward reverse paths are
\bea 
\xymatrix{\ar @{} [dr] |{\rm F}
 A,+ \ar@{<-}[d] \ar@/^1pc/ @{->} [r] &B,+ \ar@{->}[d]    & \ar @{} [dr] |{\rm R}  A,+ \ar@{->}[d]  \ar@/^1pc/ @{<-} [r] &  B,+ \ar@{<-}[d] \\
  A,-   \ar@/_1pc/ @{<-} [r]       &B,-    &A,-  \ar@/_1pc/ @{->} [r]   &B,-  } \nonumber \\
\\
\xymatrix{ \ar @{} [dr] |{\rm B} A,+ \ar@{<-}[d] \ar@/_1pc/ @{->} [r] &B,+ \ar@{->}[d] & \ar @{} [dr] |{\rm BR} A,+ \ar@{->}[d] \ar@/_1pc/ @{<-} [r] &B,+ \ar@{<-}[d]\\
 A,-   \ar@/^1pc/ @{<-} [r]       &B,-  &A,-   \ar@/^1pc/ @{->} [r]       &B,-} \nonumber
 \eea
The probability for completion of a cycle is proportional to the product of the transition constants in the cycle \cite{ast_pra89}.   The proportionality constants involve rate constants for back and forth transitions, lifetimes of the states within the cycle, etc.  Importantly, since F, R, B, and BR directional cycles (Eq. (26)) involve the same states, the proportionality constants are the same for all of these symmetry related cycles. Thus, with $k_{\rm i_\pm j_\pm} = k_{\rm ij}(\psi(t) = \pm\Psi)$, it is easy to derive  
\bea 
\frac{P_{\rm F}}{P_{\rm R}} = \frac{\roarrow{k}_{A_+B_+} \roarrow{k}_{B_-A_-}}{\loarrow{k}_{B_+A_+} \loarrow{k}_{A_-B_-}} = e^{W - \Delta U} \nonumber \\
\\
\frac{P_{\rm B}}{P_{\rm BR}}=\frac{\loarrow{k}_{A_+B_+} \loarrow{k}_{B_-A_-}}{\roarrow{k}_{B_+A_+} \roarrow{k}_{A_-B_-}} = e^{W + \Delta U} \nonumber
\eea
where $W = \ln{(\phi_+^{\rm o})}$ is the work done in the forward cycle when the energy is increased by $2 z_0 \Psi \sin{(s \pi)}$ in going from $A,- \rightarrow A,+$ and again from $B,+ \rightarrow B,-$ .  Here and below $\phi_+^{\rm (e,o)} = \phi^{\rm (e,o)}(\psi(t) = +\Psi)$.   The ratio of the probabilities for a forward and backward cycle is
\be
\frac{P_{\rm F}}{P_{\rm B}} =\frac{\roarrow{k}_{A_+B_+} \roarrow{k}_{B_-A_-}}{\loarrow{k}_{A_+B_+} \loarrow{k}_{B_-A_-}} = e^{-\Delta U} \phi_+^{\rm e}
\ee
and the ratio of the net forward to backward steps is
\be
\frac{P_{\rm F} + P_{\rm BR}}{P_{\rm B} + P_{\rm R}} =  e^{-\Delta U} \left(\frac{ 1+\phi_+^{\rm e} \phi_+^{\rm o}  }{\phi_+^{\rm o}  +\phi_+^{\rm e} }\right)
\ee
The expansion of the coefficient in Eq. (29) involves only even powers of the amplitude $\psi_0$ of the external driving - the Brownian motor mechanism is a fundamentally non-linear effect of the external driving \cite{ast_jcp89,bier_pre00}.  It does not necessarily require a large amplitude driving to observe experimentally \cite{liu_jbc90} however since the linear term is, by symmetry, identically zero.  The non-monotonic frequency response observed experimentally \cite{liu_jbc90} and explained theoretically \cite{rob_jcp91} arises from the uncoupled trajectories which take on greater or lesser importance depending on the frequency of the applied signal.   The symmetry relations derived here are evidently frequency independent. 

 Many recent synthetic implementations of molecular brownian motors involve motion between discrete binding sites.  Without an external driving the thermally activated transitions show no long time order irrespective of structural asymmetry, in consistency with the principle of detailed balance.  By using external energy to manipulate the environment, even in a seemingly random way, it is possible to break detailed balance and  to drive directed motion.  The symmetry relations derived here remain valid even under the action of an external perturbation and provide insight into how it may be possible to optimize synthetic molecular motors.


\begin{thebibliography}{99}
\bibitem{ast_sci97} R. D. Astumian, 
{\em Science} {\bf 276}, 917--922 (1997).

\bibitem{rei_pr02}
P. Reimann, {\em Phys. Rep. } {\bf 361}, 57--265  (2002).

\bibitem{ast_pt02}
R. D. Astumian and P. Hanggi, 
{\em Phys. Today} {\bf 55} (11), 33--39 (2002).

\bibitem{ast_ebj98}
R. D. Astumian and I. Derenyi,
{\em Eur. Biophys. J.} {\bf 27}, 474--489 (1998).

\bibitem{leigh_angchem07}
E.R. Kay, D.A. Leigh, F. Zerbetto,  {\em Ang. Chem. Int. Ed.}, {\bf 46}, 72-191 (2007).

\bibitem{fer_natnano06}
W.R. Browne and B.L. Feringa, {\em Nature Nanotechnology}, {\bf 1}, 25-35 (2006).

\bibitem{ons_pr53} L. Onsager and S. Machlup, 
{\em Phys. Rev.} {\bf 91}, 1505--1512 (1953). 

\bibitem{ast_pra89}
R.D. Astumian, P.B. Chock, T.Y. Tsong, and H.V. Westerhoff, {\em Phys. Rev. A}  {\bf 39}, 6416 (1989).

\bibitem{ast_prl94}
R. D. Astumian and M. Bier, {\em Phys. Rev. Lett.} {\bf 72}, 1766--1769 (1994).

\bibitem{pro_prl94}
J. Prost, L. Peliti, A. Ajdari, {\em Phys. Rev. Lett.} {\bf 72} 2652--2655 (1994).

\bibitem{boch_jetp77}
G. N. Bochkov and Yu. E. Kuzovlev, {\em Sov. Phys. JETP} {\bf 45}, 125--130 (1977).

\bibitem{boch_physa81}
G. N. Bochkov and Yu. E. Kuzovlev, 
{\em Physica A} {\bf 106}, 443--479 (1981).

\bibitem{gil_ajp95}
D. T. Gillespie,
{\em Am. J. Phys.} {\bf 64}, 225--239 (1995).

\bibitem{crooks_jstatphys98}
G. Crooks, {\em J. Stat. Phys.} {\bf 90}, 1481--1493 (1998).

\bibitem{bier_pla98}
M. Bier, R. D. Astumian, {\em Phys. Let. A} {\bf 247}, 385 (1998).

\bibitem{bier_pre99}
M. Bier, I., Derenyi, M. Kostur, and R.D. Astumian,
 {\em Phys. Rev. E.} {\bf 59}, 6422--6432 (1999).
 
\bibitem{ast_jcp89}
R.D. Astumian and B. Robertson, {\em J. Chem. Phys.} {\bf 91}, 4891--4901 (1989). 

\bibitem{sin_epl07}
N.A. Sinitsyn and I. Nemenman, {\em Europhys. Lett.} {\bf 77}, 58001 (2007).

\bibitem{bier_pre00}
M. Bier, M. Kostur, I. Derenyi, and R.D. Astumian, {\em Phys. Rev. E} {\bf 61}, 7184--7187 (2000). 

\bibitem{liu_jbc90}
D.S. Liu, R.D. Astumian, and T.Y. Tsong, {\em J. Biol. Chem.} {\bf 265}, 7260--7267 (1990).

\bibitem{rob_jcp91}
B. Robertson and R.D. Astumian, {\em J. Chem. Phys.} {\bf 94}, 7414--7419 (1991). 


\end{thebibliography}
\end{document}